\documentclass[12pt]{article}
\usepackage{color}
\usepackage[dvips]{graphicx,psfrag}
\usepackage{amsmath,amssymb}
\usepackage{bm}
\baselineskip=\normalbaselineskip
\renewcommand{\baselinestretch}{1.34}
\makeatletter
\@addtoreset{equation}{section}
\makeatother

\renewcommand{\thefootnote}{\fnsymbol{footnote}}

\newcommand{\vev}[1]{\left\langle #1 \right\rangle}
\newcommand{\ket}[1]{\left|#1\right\rangle}
\newcommand{\bra}[1]{\left\langle #1\right|}

\newcommand{\maru}[1]
{{\ooalign{\hfil#1\/\hfil\crcr\raise.167ex\hbox{\mathhexbox20D}}}}

\newcommand{\cM}{\mathcal{M}}

\newcommand{\bC}{\mathbb{C}}

\newcommand{\bZ}{\mathbb{Z}}

\newcommand{\eq}[1]{(\ref{#1})}

\newcommand{\nn}{\nonumber}

\allowdisplaybreaks[1]

\newcommand{\dphi}{\partial\varphi}
\newcommand{\cb}{\bar{c}}
\newcommand{\del}{\partial}
\newcommand{\gint}{\oint\hspace{-14pt}\bigcirc}

\newcommand{\ddz}{\frac{d\zeta}{2\pi i}}

\begin{document}
\setcounter{page}{0}
\begin{flushright}
\parbox{40mm}{%
KUNS-1971 \\
{\tt hep-th/0505253} \\
May 2005}

\end{flushright}

\vfill

\begin{center}
{\Large{\bf 
Comments on the D-instanton calculus \\
in $(p,p+1)$ minimal string theory
}}
\end{center}

\vfill

\begin{center}
{\large{Masafumi Fukuma}}\footnote%
{E-mail: {\tt fukuma@gauge.scphys.kyoto-u.ac.jp}},  
{\large{Hirotaka Irie}}\footnote%
{E-mail: {\tt irie@gauge.scphys.kyoto-u.ac.jp}} and 
{\large{Shigenori Seki}}\footnote%
{E-mail: {\tt seki@gauge.scphys.kyoto-u.ac.jp}}  \\[2em]
Department of Physics, Kyoto University, 
Kyoto 606-8502, Japan \\

\end{center}

\vfill
\renewcommand{\thefootnote}{\arabic{footnote}}
\setcounter{footnote}{0}
\addtocounter{page}{1}

\begin{center}
{\bf abstract}
\end{center}

\begin{quote}
The FZZT and ZZ branes in $(p,p+1)$  minimal string theory
are studied in terms of continuum loop equations. 
We show that systems in the presence of ZZ branes (D-instantons) 
can be easily investigated 
within the framework of the continuum string field theory 
developed by Yahikozawa and one of the present authors \cite{fy1}. 
We explicitly calculate the partition function of a single ZZ brane 
for arbitrary $p$.  
We also show that the annulus amplitudes of ZZ branes 
are correctly reproduced. 

\end{quote}
\vfill
\renewcommand{\baselinestretch}{1.4}
\newpage

\section{Introduction}

Since conformally invariant boundary states were constructed 
in the worldsheet description \cite{DOZZ,fzz-t,zz}, 
renewed interest in noncritical string theory has arisen 
\cite{mgv,Mar,kma,holodual,c1inst,cj2,SeSh,kk,Hanada:2004im,%
chemi,KOPSS,tt,tesch}.  
These boundary states lead to two types of branes.
One is of FZZT branes given by Neumann-like boundary states.
They are extended in the weak coupling region of Liouville coordinate 
and naturally correspond to unmarked macroscopic-loop
operators \cite{bdss} in matrix models.
The other is of ZZ branes given by Dirichlet-like boundary states.
They are localized in the strong coupling region of Liouville coordinate 
and are shown in \cite{mgv,Mar,kma} 
to correspond to ``eigenvalue instantons'' in matrix models \cite{david}.
In particular, the decay amplitude of ZZ branes is identified with
that of eigenvalues rolling from a local maximum of matrix-model  
potential,
and the ZZ branes are identified with D-instantons 
in noncritical string theory \cite{p-g,s,fy3}.%
\footnote{
In what follows, the terms ``ZZ branes" and ``D-instantons" 
will be used interchangeably.
}

In this paper, we show that these FZZT and ZZ branes
can be naturally understood within the framework of
a continuum string field theory for macroscopic loops \cite 
{fy1,fy2}.
The Fock space of this string field theory is realized by
$p$ pairs of free chiral fermions $c_a(\zeta)$ and $\cb_a 
(\zeta)$
$(a=0,1,\cdots,p-1)$,
living on the complex plane whose coordinate is given 
by the boundary cosmological constant $\zeta$.
It is shown in \cite{fy1} that
their diagonal bilinears $\cb_a(\zeta)c_a(\zeta)$
(bosonized as $\partial\varphi_a(\zeta)$) can be
identified with marked macroscopic loops,
\begin{align}
 \partial\varphi_a(\zeta)=\,:\!\cb_a 
  (\zeta) c_a(\zeta)\!:\,
   =\int_0^\infty\!dl\,e^{-\zeta l} 
\,\Psi(l)
\end{align}
with $\Psi(l)$ being the operator
creating the boundary of length $l$.
This implies that the unmarked macroscopic loops (FZZT branes)
are described by
\begin{align}
\varphi_a(\zeta)\sim \int_0^\infty 
\!\frac{dl}{l}\,
   e^{-\zeta l}\,\Psi(l).
\end{align}
{}Furthermore, it is shown in \cite{fy1}
that their off-diagonal bilinears $\cb_a(\zeta)c_b 
(\zeta)$ ($a\neq b$)
(bosonized as $e^{\varphi_a(\zeta)-\varphi_b 
(\zeta)}$)
can be identified with the operator creating a soliton 
at the ``spacetime coordinate" $\zeta$ \cite{fy3}.
In order for this operator to be consistent with
the continuum loop equations (or the $W_{1+\infty}$ constraints)
\cite{fkn1,dvv,g,fkn2,fkn3},
the position of the soliton must be integrated as
\begin{align}
   D_{ab}\equiv \int \frac{d\zeta}{2\pi  
i}\,\cb_a(\zeta)c_b(\zeta)
    =\int \frac{d\zeta}{2\pi i}\,e^ 
{\varphi_a(\zeta)-\varphi_b(\zeta)}.
\end{align}
This integral can be regarded as defining an effective theory
for the position of the soliton.
In the weak coupling limit $g\rightarrow 0$
($g$: the string coupling constant)
the expectation value of $\cb_a(\zeta)c_b(\zeta)$
behaves as $\exp\bigl(g^{-1}\Gamma_{ab}(\zeta) 
+O(g^0)\bigr)$, 
where the ``effective action" $\Gamma_{ab}$ is expressed 
as the difference of the disk amplitudes: 
\begin{align}
 \Gamma_{ab} = \vev 
 {\varphi_a}^{\!(0)}-\vev{\varphi_b}^{\!(0)}. 
 \label{basic-eq}
\end{align}
Thus, in the weak coupling limit, 
the soliton will get localized at a saddle point of $\Gamma_{ab}$ 
and behave as a D-instanton (a ZZ brane). 
The relation \eq{basic-eq} evaluated at the saddle point 
gives a well-known relation
between disk amplitudes of FZZT and of ZZ branes \cite{Mar}.
One shall be able to extend this relation  
to arbitrarily higher-order amplitudes.

The main aim of the present paper is to elaborate the calculation
around the saddle point performed in \cite{fy1},
and to establish the above relationship
with a catalog of possible quantum numbers.
We show that the partition functions of D-instantons 
for generic $(p,p+1)$ minimal string theories can be calculated 
explicitly, 
and also demonstrate that 
many of the known results obtained in Liouville field theory  
\cite{Mar,SeSh,KOPSS}
and/or in matrix models \cite{david,Hanada:2004im,chemi}
can be reproduced easily.

This paper is organized as follows.
In section 2 we give a brief review on the continuum string field theory
for macroscopic loops \cite{fy1,fy2}.
In section 3 we evaluate one-instanton partition function
for $(p,p+1)$ minimal string theory,
and show that it correctly reproduces 
the value obtained in \cite{david} 
(see also \cite{Hanada:2004im,chemi}) for the pure gravity case ($p=2$).
In section 4 we make a detailed comparison of our analysis 
with that performed in Liouville field theory. 
In section 5 we calculate the annulus amplitudes of ZZ branes,
and show that it reproduces the values 
obtained in \cite {Mar,KOPSS}.
Section 6 is devoted to conclusion and discussions.
In the present paper, we exclusively consider the unitary case
$(p,q)=(p,p+1)$.
General $(p,q)$ minimal strings can also be treated in a similar manner,
and detailed analysis will be reported in the forthcoming paper.

\section{Review of the noncritial string field theory}

{}From the viewpoint of noncritical strings, 
$(p,\,p+1)$ minimal string theory describes 
two-dimensional gravity 
coupled to matters of minimal unitary conformal field theory 
($c_\mathrm{matter}=1-6/p(p+1)$). 
There the basic physical operators are the macroscopic loop operator 
$\Psi(l)$ which creates a boundary of length $l$. 
It is often convenient to introduce their Laplace transforms 
\begin{align}
 \del\varphi_0(\zeta)\equiv \int_0^\infty\! dl\,e^{-\zeta l}\,\Psi(l), 
\end{align}
and $\zeta$ is called the boundary cosmological constant. 
The left-hand side is written with a derivative for later convenience. 
Note that $\zeta$ can take a different value on each boundary. 
The connected correlation functions of macroscopic loop operators 
\begin{align}
 \bigl<\del\varphi_0(\zeta_1)\cdots\del\varphi_0(\zeta_n)\bigr>_{\rm \!c}
\end{align} 
are expanded with respect to the string coupling constant $g~(>0)$ as
\begin{align}
 \bigl<\del\varphi_0(\zeta_1)\cdots\del\varphi_0(\zeta_n)\bigr>_{\rm \!c}
 =\sum_{h\geq 0}\,
  g^{-2+2h+n}\,
  \bigl<\del\varphi_0(\zeta_1)
   \cdots\del\varphi_0(\zeta_n)\bigr>^{\!\!(h)}_{\rm \!c},
\end{align}
where $h$ is the number of handles. 
Note that 
 $\bigl<\partial\varphi_0(\zeta)\bigr>^{\!\!(0)}$ 
and 
 $\bigl<\partial\varphi_0(\zeta_1)\,
    \partial\varphi_0(\zeta_2)\bigr>^{\!\!(0)}_{\rm \!c}$ 
correspond to the disk and annulus (cylinder) amplitudes, 
respectively.

The operator formalism of the continuum string field theory 
\cite{fy1} is constructed with the following steps: 

\noindent\underline{STEP1}: \\
We introduce $p$ pairs of free chiral fermions $c_a(\zeta)$ 
and $\cb_a(\zeta)$ ($a=0,1,\cdots,p-1$) which have the OPE
\begin{align}
 c_a(\zeta)\,\cb_b(\zeta')\sim \frac{\delta_{ab}}{\zeta-\zeta'}
  \sim \cb_a(\zeta)\,c_b(\zeta').
\end{align}
This can be bosonized with $p$ free chiral bosons $\varphi_a(\zeta)$ 
(defined with the OPE: 
$\varphi_a(\zeta)\,\varphi_b(\zeta')
 \sim+\,\delta_{ab}\,\ln(\zeta-\zeta')$)  
as
\begin{align}
 \cb_a(\zeta)=K_a :\!e^{\varphi_a(\zeta)}\!:,\quad
  c_a(\zeta)=K_a :\!e^{-\varphi_a(\zeta)}\!:,
\end{align}
where $K_a$ are cocycles which ensure the correct anticommutation 
relations between fermions with different indices $a\neq b$.%
\footnote{
In \cite{fy1} $K_a$ are chosen to be 
$K_a\equiv \prod_{b=0}^{a-1}(-1)^{p_b}$ 
with $p_a$ being the fermion number of the $a$-th species.
}
In this paper the normal ordering $:~:$ are always taken 
so as to respect the ${\rm SL}(2,\bC)$-invariant vacuum $\ket{0}$.
The chiral field $\varphi_a(\zeta)$ can in turn be expressed as 
\begin{align}
 \del\varphi_a(\zeta)=\,:\!\cb_a(\zeta)\,c_a(\zeta)\!:.
\end{align}
\noindent\underline{STEP2}: \\
We impose the $\bZ_p$-twisted boundary conditions on the fermions as 
\begin{align}
   c_a(e^{2\pi i}\zeta)=c_{[a+1]}(\zeta),\quad
  \cb_a(e^{2\pi i}\zeta)=\cb_{[a+1]}(\zeta)\quad
  \bigl([a]\equiv a~\mbox{(mod~$p$)}\bigr), 
\end{align}
and correspondingly, 
\begin{align}
 \del\varphi_a(e^{2\pi i}\zeta)=\del\varphi_{[a+1]}(\zeta).
\end{align}
This can be realized by inserting the $\bZ_p$-twist field 
$\sigma(\zeta)$ at $\zeta=0$ and at $\zeta=\infty$, 
with which 
the chiral field $\del\varphi_a(\zeta)$ is expanded as 
\begin{align}
 \bra{\sigma}\cdots \del\varphi_a(\zeta)\cdots\ket{\sigma}
  &=\bra{\sigma}\cdots \frac{1}{p}\sum_{n\in\bZ}
  \omega^{-na}\alpha_n\,\zeta^{-n/p-1}\cdots\ket{\sigma}, 
 \quad \bigl(\omega \equiv  e^{2\pi i /p}\bigr)\\
 &~~~~~~~\bigl[\alpha_n,\,\alpha_m\bigr]=n\,\delta_{n+m,0}.
\end{align}
Here $\bra{\sigma}\equiv\bra{0}\!\sigma(\infty)$ and 
$\ket{\sigma}\equiv\sigma(0)\!\ket{0}$.

\noindent\underline{STEP3}: \\
We introduce the generators of the $W_{1+\infty}$ algebra \cite{winf},
$\{W^k_n\}$ $(k=1,2,\cdots;\,n\in\bZ)$, 
that are given by the mode expansion of the currents 
\begin{align}
 W^k(\zeta)\,\equiv\,\sum_{n\in\bZ}W^k_n\,\zeta^{-n-k}
  \,=\, \sum_{a=0}^{p-1}
  :\!\cb_a(\zeta)\,\del_{\!\zeta}^{\,k-1}c_a(\zeta)\!:
  \quad (k=1,2,\cdots).
\end{align}
Finally we introduce the state $\ket{\Phi}$ 
that satisfies the vacuum condition of the $W_{1+\infty}$ constraints:
\begin{align}
 W^k_n\,\ket{\Phi}=0~~~(k\geq1,~n\geq -k+1).
 \label{winf_constraint}
\end{align}
This constraint is shown to be equivalent to 
the continuum loop equations \cite{fkn1,dvv,gn,g,fkn2,fkn3}. 
In addition to the $W_{1+\infty}$ constraints,
we further require that $\ket{\Phi}$ be a decomposable state.%
\footnote{
A state $\ket{\Phi}$ is said to be decomposable
if it can be written as $\ket{\Phi}=e^H \ket{\sigma}$,
where $H$ is a bilinear form of the fermions, 
$H=\displaystyle{\int}\!d\zeta\,d\zeta'\,\sum_{a,b}\cb_a(\zeta)\,
h_{ab}(\zeta,\zeta')\,c_b(\zeta')$.
This is equivalent to the statement that
$\tau(x)=\bra{\sigma} \exp\{\sum_{n=1}^\infty x_n\alpha_n\}
\ket{\Phi}$ is a $\tau$ function of the $p$-th reduced KP hierarchy 
\cite{djkm}.
It is proved in \cite{fkn2,fkn3} that
this set of conditions ($W_{1+\infty}$ constraints and
decomposability) is equivalent to
the Douglas equation, $[P,\,Q]=1$ \cite{md}.
}

\noindent\underline{STEP4}: \\
One can prove that the connected correlation functions 
of macroscopic loop operators are given as cumulants 
(or connected parts) of the following correlation functions \cite{fy1}:%
\footnote{
Since the normal ordering differs from that for the twisted vacuum 
$\ket{\sigma}$, 
the two-point function acquires a finite renormalizaion. 
This turns out to be the so-called nonuniversal term 
in the annulus amplitude \cite{fy1}.
The representation of loop correlators with free twisted bosons can also be 
found in \cite{cj1}, where open-closed string coupling is investigated.
} 
\begin{align}
 \bigl<\del\varphi_0(\zeta_1)\cdots\del\varphi_0(\zeta_n)\bigr>
  &\equiv 
  \frac{
  \left\langle\left.\left.\left.\!
  -\frac{B}{g}\,\right|\,
  :\!\dphi_0(\zeta_1)\cdots\dphi_0(\zeta_n)\!:
  \,\right|\,\Phi\,\right\rangle\right.}{
  \left\langle\left.\left.\!
  -\frac{B}{g}\,\right|\,
  \,\Phi\,\right\rangle\right.}.
\end{align}
Here the state 
\begin{align}
 \bra{-\frac{B}{g}} \equiv \Bigl<\sigma\Bigr|
  \exp\biggl(-\frac{1}{g}\sum_{n=1}^{\infty}B_n \alpha_n\biggr)
\end{align}
characterizes the theory, 
and the $(p,q)$ minimal string is realized 
by taking $B_{p+q}\neq 0$ and $B_n=0$ $(\forall n\geq p+q+1)$.

{}From the viewpoint of two-dimensional gravity, 
$\alpha_1$ creates the lowest-dimensional operator on random surfaces, 
so that $B_1~(\equiv\mu)$ should correspond 
to the bulk cosmological constant 
in the unitary minimal strings.  
In fact, if we choose 
$B_1=\mu,\,B_{2p+1}=-4p/(p+1)(2p+1)$ and $B_n=0~(n\neq 0~{\rm or}~2p+1)$, 
then we obtain the following disk and annulus amplitudes 
for marked macroscopic loops \cite{fy1}: 
\begin{align}
 \bigl<\partial\varphi_0(\zeta)\bigr>^{\!\!(0)}
  &=\frac{2^{(p-1)/p}}{p+1}\,
    \Bigl[\bigl(\zeta+\sqrt{\zeta^2-\mu}\bigr)^{(p+1)/p}
   +\bigl(\zeta-\sqrt{\zeta^2-\mu}\bigr)^{(p+1)/p}\Bigr], \label{dphi}\\
 \bigl<\partial\varphi_0(\zeta_1)\,
    \partial\varphi_0(\zeta_2)\bigr>^{\!\!(0)}_{\rm \!c}
  &=\partial_{\zeta_1}\partial_{\zeta_2}\biggl[
   \ln\biggl\{\Bigl(\zeta_1+\sqrt{\zeta_1^2-\mu}\Bigr)^{1/p}
             +\Bigl(\zeta_1-\sqrt{\zeta_1^2-\mu}\Bigr)^{1/p}\nn\\
  &~~~~~~~~~~~~~~~~~~~-\Bigl(\zeta_2+\sqrt{\zeta_2^2-\mu}\Bigr)^{1/p}
             -\Bigl(\zeta_2-\sqrt{\zeta_2^2-\mu}\Bigr)^{1/p}\biggr\}\nn\\
  &~~~~~~~~~~~~-\ln\bigl(\zeta_1-\zeta_2\bigr)\biggr], \label{dphidphi}
\end{align}  
which agree with the matrix model results \cite{bdss}.
The amplitudes of the corresponding FZZT branes 
are obtained by integrating the above amplitudes. 
In particular, the annulus amplitudes of the FZZT branes 
are given by 
\begin{align}
 \bigl<\varphi_0(\zeta_1)\,\varphi_0(\zeta_2)\bigr>^{\!\!(0)}_{\rm \!c}
  &=\ln\biggl\{\Bigl(\zeta_1+\sqrt{\zeta_1^2-\mu}\Bigr)^{1/p}
             +\Bigl(\zeta_1-\sqrt{\zeta_1^2-\mu}\Bigr)^{1/p}\nn\\
  &~~~~~~~~~~~~~~~~~~~-\Bigl(\zeta_2+\sqrt{\zeta_2^2-\mu}\Bigr)^{1/p}
             -\Bigl(\zeta_2-\sqrt{\zeta_2^2-\mu}\Bigr)^{1/p}\biggr\}\nn\\
  &~~~~~~~~~~~~-\ln\bigl(\zeta_1-\zeta_2\bigr),
 \label{phiphi}
\end{align}
with nonuniversal additive constants.

Moreover, by combining the $W_{1+\infty}$ constraint with the KP equation, 
one can easily show that the two-point function 
of cosmological term 
$u(\mu,g)\equiv \bigl(-g\,\frac{\del}{\del\mu}\bigr)^2 
\ln \bigl<-B/g\,\bigr|\Phi\bigr>$ 
satisfies the Painlev\'e-type equations \cite{bk-ds-gm, b-g-c}%
\footnote{
If one takes account of the doubling in matrix models 
with even potentials, 
the two-point function will be replaced by 
$f\equiv 2u$. 
}
\begin{align}
 (p=2)\quad& 4u^2+\frac{2g^2}{3}\,\del_\mu^2 u =\mu,\\
 (p=3)\quad& 4u^3+\frac{3g^2}{2}\,(\del_\mu u)^2
      +3g^2 u\,\del_\mu^2u +\frac{g^4}{6}\,\del_\mu^4 u =-\mu,\\
 & \vdots \nn
\end{align}

As for the solutions with soliton backgrounds, 
the crucial observation made in \cite{fy1} is that 
the commutators between the $W_{1+\infty}$ generators 
and $\cb_a(\zeta)c_b(\zeta)$ $(a\neq b)$ give total derivatives: 
\begin{align}
 \bigl[W^k_n,\,\cb_a(\zeta)c_b(\zeta)\bigr]=\partial_\zeta\bigl(*\bigr),
\end{align}
and thus the operator 
\begin{align}
 D_{ab}\equiv \gint\frac{d\zeta}{2\pi i}\,
  \cb_a(\zeta)c_b(\zeta)
 \label{D-instanton}
\end{align} 
commutes with the $W_{1+\infty}$ generators: 
\begin{align}
 \bigl[W^k_n,\,D_{ab}\bigr]=0.
 \label{commzero}
\end{align}
Here the contour integral in \eq{D-instanton} 
needs to surround the point of infinity ($\zeta=\infty$) $p$ times 
in order to resolve the $\bZ_p$ monodromy.
Equation \eq{commzero} implies that 
if $\ket{\Phi}$ is a solution of the $W_{1+\infty}$ constraints 
\eq{winf_constraint}, 
then so is $D_{a_1 b_1}\cdots D_{a_r b_r}\ket\Phi$. 
We will see that the latter can actually be identified 
with an $r$-instanton solution.%
\footnote{
Note that if the decomposability condition 
is further imposed, 
the only possible form for the collection of instanton solutions 
should be $\ket{\Phi,\theta}\equiv\prod_{a\neq b}\exp\bigl(
\theta_{ab}\,D_{ab}\bigr)\ket\Phi$ 
with fugacity $\theta_{ab}$ \cite{fy2}.
}

By using the weak field expansions, 
the expectation value of $D_{ab}$ can be expressed as
\begin{align}
 \vev{D_{ab}}&=\gint\ddz \vev{e^{\varphi_a(\zeta)-\varphi_b(\zeta)}}\nn\\
  &=\gint\ddz 
  \exp\left\{\vev{e^{\varphi_a(\zeta)-\varphi_b(\zeta)}-1}_{\rm \!c}
  \right\}\nn\\
 &=\gint\ddz \exp\left\{
  \bigl<\varphi_a(\zeta)-\varphi_b(\zeta)\bigr>
  +\frac{1}{2}\bigl<\left(
      \varphi_a(\zeta)-\varphi_b(\zeta)
   \right)^2\bigr>_{\rm \!c}
  +\,\cdots\right\}.
 \label{Dab1}
\end{align}
Since connected $n$-point functions have the following expansion in $g$: 
\begin{align}
 \bigl<\partial\varphi_{a_1}(\zeta_1)
  \cdots\partial\varphi_{a_n}(\zeta_n)\bigr>_{\rm \!c}
  =\sum_{h=0}^\infty \,g^{-2+2h+n}\,
  \bigl<\partial\varphi_{a_1}(\zeta_1)\cdots
   \partial\varphi_{a_n}(\zeta_n)\bigr>_{\rm \!c}^{\!\!(h)} ,
\end{align}
leading contributions to the exponent of \eq{Dab1} 
in the weak coupling limit 
come from spherical topology ($h=0$):
\begin{align}
  \vev{D_{ab}}&=\gint\ddz\,
    e^{\,(1/g)\,\Gamma_{ab}(\zeta)
    \,+\,(1/2)\,K_{ab}(\zeta) \,+\,O(g)}
 \label{integral}
\end{align}
with 
\begin{align}
 \Gamma_{ab}(\zeta)\,\equiv\,\bigl<\varphi_a(\zeta)\bigr>^{\!\!(0)}
   -\bigl<\varphi_b(\zeta)\bigr>^{\!\!(0)}, \qquad
 K_{ab}(\zeta)\,\equiv\,\bigl<\bigl(\varphi_a(\zeta)-
   \varphi_b(\zeta)\bigr)^2\bigr>_{\rm \!c}^{\!\!(0)}.
\end{align}
Thus, in the weak coupling limit 
the integration is dominated by the value around a saddle point  
in the complex $\zeta$ plane.

\section{Saddle point analysis}

In this section, 
we make a detailed calculation of the integral \eq{integral} 
around a saddle point, up to $e^{\,O(g^1)}$.

The functions $\Gamma_{ab}(\zeta)$ and $K_{ab}(\zeta)$ 
can be calculated, 
basically by integrating the disk and annulus amplitudes 
(\eq{dphi} and \eq{dphidphi}), 
followed by analytic continuation 
$\zeta_1\rightarrow e^{2\pi ia}\zeta_1$, 
$\zeta_2\rightarrow e^{2\pi i b}\zeta_2$ 
and by taking the limit $\zeta_1,\,\zeta_2\rightarrow \zeta$. 
For example, $\Gamma_{ab}$ is calculated as
\begin{align}
 \Gamma_{ab}(\zeta)&=
  \bigl<\varphi_a(\zeta)\bigr>^{\!\!(0)}
   -\bigl<\varphi_b(\zeta)\bigr>^{\!\!(0)}\nn\\
  &=\bigl<\varphi_0({e}^{2\pi ia}\zeta)\bigr>^{\!\!(0)}
  -\bigl<\varphi_0({e}^{2\pi ib}\zeta)\bigr>^{\!\!(0)}\nn\\
 &=\int_{e^{2\pi ib}\zeta}^{e^{2\pi ia}\zeta}\! 
  d\zeta'\,\bigl<\partial\varphi_0(\zeta')\bigr>^{\!\!(0)},
\end{align}
and we obtain
\begin{align}
 \Gamma_{ab}(\zeta)&=\frac{p}{2^{1/p}(p+1)}\,\mu^{(2p+1)/2p}\,
  \,\left[\, \frac{1}{2p+1}\Bigl\{
  (\omega^a-\omega^b)s^{(2p+1)/p}+
  (\omega^{-a}-\omega^{-b})s^{-(2p+1)/p}\Bigr\}
  \right.\nn\\
 &~~~~~~-\,\left.\Bigl\{
  (\omega^a-\omega^b)s^{1/p}+
  (\omega^{-a}-\omega^{-b})s^{-1/p}\Bigr\}
  \,\right],
 \label{Zzzab} 
\end{align}
where $\omega\equiv e^{2\pi i/p}$, 
and $s=s(\zeta)$ is defined as 
\begin{align}
 s=\frac{\zeta}{\sqrt{\mu}} 
   + \sqrt{\biggl(\frac{\zeta}{\sqrt{\mu}}\biggr)^2-1}
  ,\qquad
 s^{-1}=\frac{\zeta}{\sqrt{\mu}} 
   - \sqrt{\biggl(\frac{\zeta}{\sqrt{\mu}}\biggr)^2-1}.
\end{align}
On the other hand, 
the calculation of $K_{ab}$ needs a special care 
because  $\vev{\varphi_a(\zeta_1)\varphi_b(\zeta_2)}$ 
does not obey simple monodromy.  
This is due to the fact that the two-point function 
 $\vev{\varphi_a(\zeta_1)\varphi_b(\zeta_2)}$ 
is defined with the normal ordering $:\ \ :$ 
that respects the ${\rm SL}(2,\mathbb{C})$ invariant vacuum:
\begin{align}
 \vev{\varphi_a(\zeta_1)\varphi_b(\zeta_2)}
 &=\frac{\vev{-B/g|:\!\varphi_a(\zeta_1)\varphi_b(\zeta_2)\!:|\Phi}}
    {\vev{-B/g|\Phi}}.
\end{align}
In fact, by using the definition 
 $:\!\varphi_a(\zeta_1)\,\varphi_b(\zeta_2)\!:\, =
 \varphi_a(\zeta_1)\,\varphi_b(\zeta_2)-\delta_{ab}\,\ln(\zeta_1-\zeta_2)$, 
the two-point functions are expressed as 
\begin{align}
 \vev{\varphi_a(\zeta_1)\varphi_b(\zeta_2)}
  &=\frac{\vev{-B/g|\,\varphi_a(\zeta_1)\varphi_b(\zeta_2)\,|\Phi}}
    {\vev{-B/g|\Phi}}
   -\delta_{ab}\ln(\zeta_1-\zeta_2)\nn\\
  &=\frac{\vev{-B/g|\,
     \varphi_0(e^{2\pi ia}\zeta_1)\varphi_0(e^{2\pi ib}\zeta_2)\,
     |\Phi}}{\vev{-B/g|\Phi}}
   -\delta_{ab}\ln(\zeta_1-\zeta_2)\nn\\
  &=\frac{\vev{-B/g|
     :\!\varphi_0(e^{2\pi ia}\zeta_1)\varphi_0(e^{2\pi ib}\zeta_2)\!:
     |\Phi}}{\vev{-B/g|\Phi}} \nn\\
  &~~~~~~~
   +\ln(e^{2\pi ia}\zeta_1-e^{2\pi ib}\zeta_2)
    -\delta_{ab}\ln(\zeta_1-\zeta_2)\nn\\
  &=\vev{\varphi_0(e^{2\pi ia}\zeta_1)\,\varphi_0(e^{2\pi ib}\zeta_2)}
   +\ln(e^{2\pi ia}\zeta_1-e^{2\pi ib}\zeta_2)
    -\delta_{ab}\ln(\zeta_1-\zeta_2).
\end{align}
We thus obtain%
\footnote{Although these amplitudes may have nonuniversal 
additive corrections, 
they will be totally canceled 
in the calculation of $\Gamma_{ab}$ and $K_{ab}$. 
In this sense, the value of the partition function of a D-instanton 
must be universal as in the matrix model cases \cite{Hanada:2004im}.
}
\begin{align}
 \bigl<\varphi_a(\zeta_1)\varphi_b(\zeta_2)\bigr>_{\rm \!c}^{\!\!(0)}
 &=\ln\biggl[\omega^a\bigl(\zeta_1+\sqrt{\zeta_1^2-\mu}\bigr)^{1/p}
  +\omega^{-a}\bigl(\zeta_1-\sqrt{\zeta_1^2-\mu}\bigr)^{1/p}\nn\\
 &~~~~~~~~-\omega^b\bigl(\zeta_2+\sqrt{\zeta_2^2-\mu}\bigr)^{1/p}
  -\omega^{-b}\bigl(\zeta_2-\sqrt{\zeta_2^2-\mu}\bigr)^{1/p}\biggr]\nn\\
 &~~~~-\delta_{ab}\ln(\zeta_1-\zeta_2),
 \label{Kab_zeta1}
\end{align}
from which $K_{ab}(\zeta)$ are calculated to be
\begin{align}
 K_{ab}(\zeta) &=\ln\Bigl(\,
   \omega^a\,s^{1/p}-\omega^{-a}\,s^{-1/p}\Bigr)
  +\ln\Bigl(\,
   \omega^b\,s^{1/p}-\omega^{-b}\,s^{-1/p}\Bigr)
  -2\ln\bigl(s-s^{-1}\bigr)\nn\\
 &~~-\ln\Bigl[\bigl(\omega^a-\omega^b)\,s^{1/p}
      +\bigl(\omega^{-a}-\omega^{-b}\bigr)\,s^{-1/p}\Bigr]\nn\\ 
 &~~-\ln\Bigl[\bigl(\omega^b-\omega^a)\,s^{1/p}
      +\bigl(\omega^{-b}-\omega^{-a}\bigr)\,s^{-1/p}\Bigr]\nn\\
 &~~+2\ln\frac{2}{p\sqrt{\mu}}.
 \label{Kab_zeta2}
\end{align}

In order to make a further calculation in a well-defined manner,  
it is convenient to introduce new variables $z$ and $\tau$ \cite{SeSh}, 
for which the functions $\Gamma_{ab}$ and $K_{ab}$ are single-valued:
\begin{align}
 s=\frac{\zeta}{\sqrt{\mu}} 
   + \sqrt{\biggl(\frac{\zeta}{\sqrt{\mu}}\biggr)^2-1}
  \equiv e^{ip\tau}, 
  \qquad 
 z\equiv \cos\tau=\frac{1}{2}(s^{1/p}+s^{-1/p}). 
 \label{param1}
\end{align}
$\zeta$ is then expressed as
\begin{align}
 \frac{\zeta}{\sqrt{\mu}}=\cos p\tau=T_p(z),
 \label{param2}
\end{align}
where $T_n(z)$ $(n=0,1,2,\cdots)$ are first Tchebycheff polynomials 
of degree $n$ defined by 
$T_n(\cos\tau)\equiv \cos n\tau$. 
The monodromy of the operator $\phi_a(z)\equiv\varphi_a(\zeta) $ 
under twisted vacuum $\ket{\sigma}$ is expressed as
\begin{align}
 \phi_a(z)\ket{\sigma}=\phi_0(z_a)\ket{\sigma}
\end{align}
with $z_a\equiv\cos\tau_a\equiv\cos(\tau+2\pi a/p)$.

We return to the calculation of the partition function 
in the presence of one soliton, $\bigl<D_{ab}\bigr>$. 
Since the measure is written as 
$d\zeta=p\sqrt{\mu}\,U_{p-1}(z)\,dz$, 
we need to calculate the following integral: 
\begin{align}
 \bigl< D_{ab} \bigr>
  =\frac{p\sqrt{\mu}}{2\pi i}\,\int\!dz\,U_{p-1}(z)\,
   e^{-(1/g)\,\Gamma_{ab}(z)+(1/2) K_{ab}(z)+O(g)}. 
\end{align}
Here $U_{n}(z)$ $(n=0,1,2,\cdots)$ are second Tchebycheff polynomials 
of degree $n$ defined by $U_n(\cos\tau)\equiv \sin(n+1)\tau/\sin\tau$. 
$T_n(z)$ and $U_n(z)$ are related as 
\begin{align}
 T_n'(z)=n\,U_n(z),\qquad
 U_n'(z)=\frac{1}{1-z^2}\,\bigl[ z\,U_n(z)-(n+1)\,T_{n+1}(z)\bigr].
\end{align}

The function $\Gamma_{ab}$ and their derivatives 
are easily obtained 
by using the formulas
\begin{align}
 \frac{dz_a}{dz}=\frac{\sin(\tau+2\pi a/p)}{\sin\tau},\qquad
 \frac{d^2 z_a}{dz^2}=\frac{\sin(2\pi a/p)}{\sin^3\tau}, 
\end{align}
and are found to be
\begin{align}
 \Gamma_{ab}(z)
  &=\bigl<\phi_a(z)\bigr>^{\!\!(0)}
    -\bigl<\phi_b(z)\bigr>^{\!\!(0)}\nn\\
  &=\frac{2^{(p-1)/p}\,p}{p+1}\mu^{(2p+1)/2p} 
   \biggl[\frac{1}{2p+1}\bigl(T_{2p+1}(z_a)-T_{2p+1}(z_b)\bigr)
  -(z_a-z_b)\biggr],
 \label{Gab}\\
 \Gamma_{ab}'(z)&=
  -\frac{2^{(2p+1)/2p}p}{p+1}\mu^{(2p+1)/2p}
  (\omega^b-\omega^a)U_{p-1}(z)y^{-p-1}(y^{2(p+1)}-\omega^{-a-b})
  \quad(y\equiv e^{i\tau}),
 \label{Gab1}\\
 \Gamma_{ab}''(z)&=\frac{z}{1-z^2}\,\Gamma_{ab}'(z) \nn\\
  &~~-\frac{p\, \omega^{(p-1)/p}}{p+1}\mu^{(2p+1)/2p}\, 
  \frac{1}{1-z^2}
   \biggl[(2p+1)\bigl(T_{2p+1}(z_a)-T_{2p+1}(z_b)\bigr)-(z_a-z_b)\biggr]. 
 \label{Gab2}
\end{align}

As for $K_{ab}$, one obtains the following formula 
from  \eq{Kab_zeta1}: 
\begin{align}
 &\bigl<\bigl(\phi_a(z_1)-\phi_b(z_1)\bigr)\,
  \bigl(\phi_c(z_2)-\phi_d(z_2)\bigr)\bigr>_{\rm \!c}^{\!\!(0)}\nn\\
 & ~~= \ln \frac{(z_{1a}-z_{2c})(z_{1b}-z_{2d})} 
             {(z_{1a}-z_{2d})(z_{1b}-z_{2c})} 
    -(\delta_{ac}+\delta_{bd}-\delta_{ad}-\delta_{bc})
      \ln\Bigl[ \sqrt{\mu}\,\bigl(T_p(z_1)-T_p(z_2)\bigr)\Bigr],
 \label{FZZT-annulus}
\end{align}
and thus $K_{ab}$ is expressed with  $z$  as
\begin{align}
 K_{ab}(z)
  &=\lim_{z_1,z_2\to z} \bigl<\bigl(\phi_a(z_1)-\phi_b(z_1)\bigr)\,
   \bigl(\phi_a(z_2)-\phi_b(z_2)\bigr)\bigr>_{\rm \!c}^{\!\!(0)}\nn\\
  &=\lim_{z_1,z_2\to z} \ln \frac{(z_{1a}-z_{2a})(z_{1b}-z_{2b})} 
             {(z_{1a}-z_{2b})(z_{1b}-z_{2a})(T_p(z_1)-T_p(z_2))^2\mu}\nn\\
  &=-\ln \bigl[-(z_a-z_b)^2U_{p-1}(z_a)U_{p-1}(z_b)\bigr]-2\ln p\sqrt{\mu}.
 \label{Kab}
\end{align}

The saddle points $z=z_*=\cos\tau_*$ are determined 
by solving the equation $\Gamma'_{ab}(z_*)=0$, 
and are found to be%
\footnote{There are other possible saddle points 
determined by $U_{p-1}(z_*)=0$. 
However, they only give irrelevant contributions to the integral 
because of the vanishing measure 
$d\zeta=p\sqrt{\mu}\,U_{p-1}(z)\,dz$ at such saddle points.
}
\begin{align}
 \tau_* = \Bigl(-\frac{a+b}{p} + \frac{a+b+l}{p+1}\Bigr)\pi, 
  \quad l \in \bZ . \label{z0a}
\end{align}
They acquire the following changes  
under the transformation $z_* \to z_{*a}=\cos(\tau_*+2\pi ia/p)$: 
\begin{align}
 z_{*a} =& \cos\left(\frac{b-a}{p}-\frac{a+b+l}{p+1}\right)\pi 
  = \cos\left(\frac{m}{p}-\frac{n}{p+1}\right)\pi,
 \label{z_a}\\
 z_{*b} =& \cos\left(\frac{b-a}{p}+\frac{a+b+l}{p+1}\right)\pi 
  = \cos\left(\frac{m}{p}+\frac{n}{p+1}\right)\pi . 
 \label{z_b}
\end{align}
Here we have introduced two integers $m$ and $n$ as 
\begin{align}
 m\equiv b-a,\qquad n\equiv a+b+l.
 \label{int_mn}
\end{align}
It is easy to see that 
$T_{2p+1}(z)$ takes the following values at those shifted points: 
\begin{align}
 T_{2p+1}(z_{*a}) =& \cos\left(\frac{m}{p}+\frac{n}{p+1}\right)\pi,\\\
 T_{2p+1}(z_{*b}) =& \cos\left(\frac{m}{p}-\frac{n}{p+1}\right)\pi . 
\end{align}
{}From this, one easily obtains 
\begin{align}
 \Gamma_{ab}(z_*)=-\frac{8p}{2^{1/p} (2p+1)}
  \,\mu^{(2p+1)/2p}
  \sin\left(\frac{n}{p+1}\pi\right)\,
  \sin\left(\frac{m}{p}\pi\right) , \label{Gamma0}
\end{align}
and
\begin{align}
 \Gamma_{ab}''(z_*)=\frac{8p}{2^{1/p} \sin^2\tau_*}
  \,\mu^{(2p+1)/2p}
  \sin\left(\frac{n}{p+1}\pi\right)\,
  \sin\left(\frac{m}{p}\pi\right)
 \label{Gamma2}.
\end{align}
By using the relation $U_{p-1}(z_{*a})
=(\sin p\tau_*/\sin \tau_{*a})\,U_{p-1}(z_*)$, 
$K_{ab}(z_*)$ can also be calculated easily, 
and is found to be
\begin{align}
K_{ab}(z_*) = 2\ln \biggl[\frac{\sqrt{
   \cos\bigl(\frac{2n\pi}{p+1}\bigr)-\cos\bigl(\frac{2m\pi}{p}\bigr)}}
 {2p\sqrt{2\mu}\,\sin\tau_*\,U_{p-1}(z_*)\,
  \sin\bigl(\frac{n\pi}{p+1}\bigr)\,
  \sin\bigl(\frac{m\pi}{p}\bigr)}\biggr].
\end{align}

In order for the integration to give such nonperturbative effects 
that vanish in the limit $g\rightarrow +0$, 
we need to choose a contour 
along which ${\rm Re\,}\Gamma_{ab}(z)$ takes only negative values.
In particular, $(m,n)$ should be chosen 
such that $\Gamma_{ab}(z_*)$ is negative. 
This in turn implies that $\Gamma_{ab}''(z_*)$ is positive, 
and thus the corresponding steepest descent path 
intersects the saddle point in the pure-imaginary direction 
in the complex $z$ plane. 
We thus take $z=z_*+it$ around the saddle point, 
so that the Gaussian integral becomes 
\begin{align}
 \vev{D_{ab}}
  &=\frac{p\sqrt{\mu}}{2\pi}\,U_{p-1}(z_*)\,
  e^{(1/2)K_{ab}(z_*)}\,e^{(1/g)\Gamma_{ab}(z_*)}\,
  \int_{-\infty}^\infty\!dt\,e^{-(1/2g)\Gamma_{ab}''(z_*)\,t^2}\nn\\
 &= p\sqrt{\frac{\mu g}{2\pi }}\,
   \frac{U_{p-1}(z_*)}{\sqrt{\Gamma_{ab}''(z_*)}}
   \, e^{(1/2)K_{ab}(z_*)}\,e^{(1/g)\Gamma_{ab}(z_*)}.
\end{align}
Substituting all the values obtained above, 
we finally get
\begin{align}
\vev{D_{ab}} &=\frac{2^{1/2p}}{8\sqrt{2\pi p}}\sqrt{g}\, 
    \mu^{-(2p+1)/4p}
     \,\sqrt{
       \frac{\cos\bigl(\frac{2n \pi}{p+1}\bigr) 
              - \cos\bigl(\frac{2m\pi}{p}\bigr)}
            {\sin^3\bigl(\frac{n\pi}{p+1}\bigr) 
             \sin^3\bigl(\frac{m\pi}{p}\bigr)}}
     \exp\left(-\frac{1}{g}\Gamma_{ba}(z_*)\right) \label{Dab0}
\end{align}
with 
\begin{align}
 \Gamma_{ba}(z_*)&=-\Gamma_{ab}(z_*) \nn\\
 &=+\,\frac{8p}{2^{1/p}\, (2p+1)}
  \,\mu^{(2p+1)/2p}
  \sin\left(\frac{n}{p+1}\pi\right)\,
  \sin\left(\frac{m}{p}\pi\right) ~(>0).
\end{align}
Note that the expression \eq{Dab0} 
is invariant under the change of $(m,n)$ into $(p-m-1,p-n)$.
Thus we can always restrict the values of $(m,n)$ 
to the region 
\begin{align}
 0<m<p-1,\quad 0<n<p, \quad m(p+1)-np>0.
\end{align}

{}For example, in the pure gravity case ($p=2$), 
the only possible choice is $(m,n)=(1,1)$ or $(a,b;l)=(0,1;0)$,  
for which we obtain 
\begin{align}
 \vev{D_{01}}=\frac{2^{1/4}\,g^{1/2}}{8\,\pi^{1/2}\,3^{3/4}}\,\mu^{-5/8}\,
  \exp\Bigl[-\frac{4\sqrt{6}}{5g}\,\mu^{5/4}\Bigr].
\end{align}
By rescaling the string coupling constant as $g=g_s/\sqrt{2}$, 
it becomes%
\footnote{
Such relations among various parameters can be best read off 
by looking at the string equations. 
}
\begin{align}
 \vev{D_{01}}=\frac{g_s^{1/2}}{8\,\pi^{1/2}\,3^{3/4}}\,\mu^{-5/8}\,
  \exp\Bigl[-\frac{8\sqrt{3}}{5g_s}\,\mu^{5/4}\Bigr].
\end{align}
This coincides, up to a factor of $i$, 
with the partition function of a D-instanton 
evaluated by resorting to one-matrix model 
\cite{david} (see also \cite{Hanada:2004im,chemi}).
We shall make a comment on this discrepancy in section 6.

\section{Comparison with Liouville field theory}

In our analysis made in the preceding sections, 
the operators which are physically meaningful 
are the macroscopic loop operator $\varphi_0(\zeta)$ 
and the soliton operators 
$\displaystyle D_{ab}
=\gint d\zeta\,\bar c_a(\zeta)\,c_b(\zeta)$ $(a\neq b)$. 
They should have their own correspondents in Liouville field theory. 
The former evidently corresponds to FZZT branes. 
For example, with the parametrization \eq{param1} and \eq{param2}, 
the annulus amplitude of FZZT branes (eq.\ \eq{phiphi}) 
is expressed (up to nonuniversal additive constants) as 
\begin{align}
 \bigl<\varphi_0(\zeta_1)\,\varphi_0(\zeta_2)\bigr>^{\!\!(0)}_{\rm \!c}
  =\ln\,\frac{z_1-z_2}{T_p(z_1)-T_p(z_2)}, 
\end{align}
and agrees with the calculation based on Liouville field theory 
\cite{Mar,KOPSS}.

On the other hand, 
the relation between the soliton operators and ZZ branes is indirect. 
In fact, our solitons can take arbitrary positions 
for finite values of $g$, 
but in the weak coupling limit 
they get localized at saddle points and become ZZ branes. 
In this section we shall establish this relationship 
between the localized solitons and the ZZ branes 
with explicit correspondence between their quantum numbers.

A detailed analysis of FZZT and ZZ branes 
in $(p,q)$ minimal string theory 
is performed in \cite{SeSh}. 
According to this, the BRST equivalence classes of ZZ branes 
are labeled by two quantum numbers $(m,n)$, 
and their boundary states can be written 
as differences of two FZZT boundary states:
\begin{align}
 \bigl|m,n\bigr>_{\rm ZZ} = \bigl|\zeta (z_{mn}^+)\bigr>_{\rm FZZT} 
  - \bigl|\zeta (z_{mn}^-)\bigr>_{\rm FZZT}.
\end{align}
Here $\zeta(z)=\sqrt{\mu}\,T_p(z)$ denotes 
the boundary cosmological constant  of an FZZT brane. 
$z_{mn}^\pm$ are the singular points 
of the Riemann surface ${\cal M}_{p,q}$ which $z$ uniformizes, 
and are given by
\begin{align}
 z_{mn}^\pm &= \cos\frac{\pi(mq\pm np)}{pq} , 
  \label{singpts} \\
  m=1,\cdots,p-1,& \quad n=1,\cdots,q-1,\quad mq-np>0.  \label{mndom}
\end{align}
{}From this, one obtains the relation 
\begin{align}
 Z_{\rm ZZ}^{(m,n)} = Z_{\rm FZZT}\bigl(\zeta(z_{mn}^+)\bigr)
  -Z_{\rm FZZT}\bigl(\zeta(z_{mn}^-)\bigr), \label{ZZFZZ}
\end{align}
where $Z_{\rm ZZ}^{(m,n)}$ is the disk amplitude of a ZZ brane 
with quantum number $(m,n)$, 
and $Z_{\rm FZZT}(\zeta)$ is that of an FZZT brane 
with boundary cosmological constant $\zeta$.

On the other hand, our analysis shows that 
in the weak coupling limit $g \to +0$,  
the partition function in the presence of a soliton, $\bigl<D_{ab}\bigr>$, 
is dominated by a saddle point $z_*$ of the function 
$\Gamma_{ab}(z)=\bigl<\phi_a(z)\bigr>^{\!\!(0)}
 -\bigl<\phi_b(z)\bigr>^{\!\!(0)}$, 
and is expressed as $\bigl<D_{ab}\bigr>\sim e^{(1/g)\Gamma_{ab}(z_*)}$.
This implies that $\Gamma_{ab}(z_*)~(<0)$ 
should be regarded as the disk amplitude of a D-instanton \cite{p-g}.
{}Furthermore, it was explicitly evaluated 
at the saddle point $z=z_*(a,b;l)
=\cos \displaystyle\Bigl(-\frac{a+b}{p}+\frac{a+b+l}{p+1}\Bigr)\pi$ 
in the previous section as 
\begin{align}
 \Gamma_{ab} (z_*) &= \bigl<\phi_a(z_*)\bigr>^{\!\!(0)} 
                - \bigl<\phi_b(z_*)\bigr>^{\!\!(0)} \nn \\
 &= \Bigl<\phi_0\bigl(z_{*a}(a,b;l)\bigr)\Bigr>^{\!\!(0)} 
          -\Bigl<\phi_0\bigl(z_{*b}(a,b;l)\bigr)\Bigr>^{\!\!(0)}\nn\\
 &= \Bigl<\varphi_0\bigl(\zeta(z_{*a}(a,b;l))\bigr)\Bigr>^{\!\!(0)} 
     -\Bigl<\varphi_0\bigl(\zeta(z_{*b}(a,b;l))\bigr)\Bigr>^{\!\!(0)} \nn\\
 &=Z_{\rm FZZT}\bigl(\zeta(z_{*a}(a,b;l))\bigr)
   -Z_{\rm FZZT}\bigl(\zeta(z_{*b}(a,b;l))\bigr). 
 \label{Gamma3}
\end{align}
Here $z_{*a}$ and $z_{*b}$ are calculated in \eq{z_a} and \eq{z_b} as
\begin{align}
 z_{*a}(a,b;l) = \cos\left(\frac{m}{p}-\frac{n}{p+1}\right)\pi,
  \quad 
 z_{*b}(a,b;l) = \cos\left(\frac{m}{p}+\frac{n}{p+1}\right)\pi, 
\end{align}
with $(a,b;l)$ being related to $(m,n)$ as in \eq{int_mn}.
We thus see that the shifted points $z_{*a}(a,b;l)$ and $z_{*b}(a,b;l)$  
correspond to the singular points $z_{mn}^\pm$ as
\begin{align}
 z_{*a}(a,b;l) = z_{mn}^-,\quad z_{*b}(a,b;l) = z_{mn}^+ .
 \label{zsdzpm}
\end{align}
Then \eq{ZZFZZ}, \eq{Gamma3} and \eq{zsdzpm} lead to 
the following equality between $\Gamma_{ab}(z_*)$ 
and $Z_{\rm ZZ}^{(m,n)}$: 
\begin{align}
 \Gamma_{ab}(z_*) = \bigl<\phi_0(z_{mn}^-)\bigr> 
   - \bigl<\phi_0(z_{mn}^+)\bigr> 
  = -\,Z_{\rm ZZ}^{(m=b-a,\,n=a+b+l)} .
 \label{rel}
\end{align}
Therefore, each saddle point corresponds to a ZZ brane.%
\footnote{
We should stress that it is not the saddle point $z_*$ 
(the ``position" of the D instanton)
but the shifted points $z_{*a}$ and $z_{*b}$ 
which actually correspond to the singular points 
in the Riemann surface ${\cal M}_{p,p+1}$, 
although the set of the saddle points 
coincides with that of the singular points.  
}
The relative minus sign between $\Gamma_{ab}(z_*)$ 
(taken to be negative for the convergence in the weak coupling limit) 
and $Z_{\rm ZZ}^{(m,n)}$ 
(conventionally normalized to be positive) 
appearing in \eq{rel} is naturally derived in our analysis 
and matched with the argument given in \cite{KOPSS}.
\\

For the rest of this section, 
we illustrate the above correspondence 
along the line of the geometric setting  
introduced in \cite{SeSh}. 
We only consider the pure gravity case $(p=2)$, 
but the generalization to other cases must be straightforward.

{}For $p=2$, the amplitude of a ZZ brane is given by $\Gamma_{10}$. 
{}Furthermore, using the $W^k_n$ constraint with $k=1$, 
one can easily show that 
 $\bigl<\del\varphi_1(\zeta)\bigr>=-\bigl<\del\varphi_0(\zeta)\bigr>$. 
Thus the saddle points of $\Gamma_{10}$ can be determined 
simply by solving the following equation:  
\begin{align}
 0 = \del_\zeta \Gamma_{01}(\zeta)
  =2\bigl<\del\varphi_0(\zeta)\bigr> = \frac{8}{3}\sqrt{f_3(\zeta)} , 
  \qquad 
  f_3(\zeta) \equiv \left(\zeta-\frac{\sqrt{\mu}}{2}\right)^2 
	 \left(\zeta + \sqrt{\mu}\right) .
\end{align}
Here $f_3(\zeta)$ is a degree-three polynomial of $\zeta$, 
and has a single root at $\zeta = -\sqrt{\mu}$ and a double root 
at $\zeta = \sqrt{\mu}/2$. 
The algebraic curve defined by $y^2 = f_3(\zeta)$ is thus a  torus 
with a pinched cycle, as depicted in Fig.~1.%
\begin{center}
\includegraphics{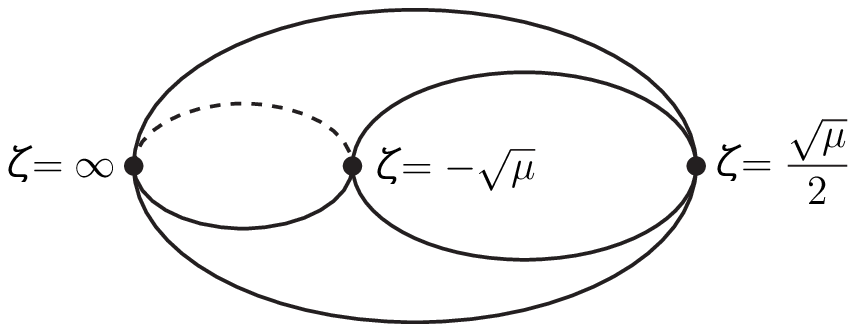} \\
\footnotesize{Fig.~1. Pinched Riemann surface} 
\end{center}
Using the relation $\zeta / \sqrt{\mu} = T_2(z)= 2z^2 - 1$, 
these roots are expressed in terms of $z$ as
\begin{align}
 \zeta_*=-\sqrt{\mu} \Longleftrightarrow z_*=0, \qquad
  \zeta_*=\frac{\sqrt{\mu}}{2} 
   \Longleftrightarrow z_*=\pm\frac{\sqrt{3}}{2} .
\end{align}
The first saddle point $z_*=0$ corresponds 
to the zero of $U_{p-1}(z)=U_1(z)=2z$ 
which was discarded in our analysis 
because such saddle points give rise to vanishing measure. 
We thus see that the pinched cycle corresponds to 
the D-instanton and thus to the ZZ brane.

If we take $(a,b;l)=(0,1;0)$ as before, 
this selects the saddle point 
at $z_*(0,1;0)=\sqrt{3}/2$. 
Then its shifted points are calculated as
\begin{align}
 z_{*0}=\frac{\sqrt{3}}{2}=z_{11}^-,\qquad
  z_{*1}=-\frac{\sqrt{3}}{2}=z_{11}^+. 
\end{align}

\section{Annulus amplitudes of D-instantons}


The annulus amplitudes of distinct D-instantons (ZZ branes) 
can also be calculated easily.
These amplitudes correspond to the states
\begin{align}
 D_{ab}\,D_{cd}\,\ket{\Phi}, 
 \label{2inst}
\end{align}
which appear, for example, 
when two distinct solitons are present in the background:
$ e^{D_{ab}+D_{cd}}\ket{\Phi}$.

The two-point function of two solitons, 
$\bigl< D_{ab}\,D_{cd} \bigr>$, 
can be written as 
\begin{align}
 \bigl< D_{ab}\,D_{cd} \bigr>
  &=\gint d\zeta \gint d\zeta'\,
   \frac{
    \bra{-B/g\,}:\!e^{\varphi_a(\zeta)-\varphi_b(\zeta)}\!:\,
   :\!e^{\varphi_c(\zeta')-\varphi_d(\zeta')}\!:\ket{\Phi} 
   }{\bigl<-B/g\,\bigr|\Phi\bigr>} \nn\\
 &=\gint d\zeta \gint d\zeta'\,
   e^{(\delta_{ac}+\delta_{bd}-\delta_{ad}-\delta_{bc})\,
     \ln(\zeta-\zeta') }\,
    \vev{e^{\varphi_a(\zeta)-\varphi_b(\zeta) 
    +\varphi_c(\zeta')-\varphi_d(\zeta')} }\nn\\
 &=\gint d\zeta \gint d\zeta'\,
   e^{(\delta_{ac}+\delta_{bd}-\delta_{ad}-\delta_{bc})\,
     \ln(\zeta-\zeta') }\,
    \exp\,\bigl< e^{\varphi_a(\zeta)-\varphi_b(\zeta) 
    +\varphi_c(\zeta')-\varphi_d(\zeta')}-1\bigr>_{\rm \!c}\nn\\
 &=\gint d\zeta \gint d\zeta'\,
   e^{(1/g)\Gamma_{ab}(\zeta)+(1/g)\Gamma_{cd}(\zeta')}\,
   e^{(1/2)K_{ab}(\zeta)}\,e^{(1/2)K_{cd}(\zeta')}\cdot \nn\\ 
 &~~~~~~\cdot e^{(\delta_{ac}+\delta_{bd}-\delta_{ad}-\delta_{bc})\,
     \ln(\zeta-\zeta') }\,
    e^{\vev{ (\varphi_a(\zeta)-\varphi_b(\zeta))\, 
     (\varphi_c(\zeta')-\varphi_d(\zeta') )}_{\rm \!c}^{\!(0)}} \, e^{O(g)}. 
\end{align}
Since $D_{ab}$ and $D_{cd}$ may have their own saddle points 
$\zeta_*$ and $\zeta_*'$ in the weak coupling limit, 
the two-point function will take the following form:
\begin{align}
 &\bigl< D_{ab}\,D_{cd} \bigr>\nn\\
 &=\bigl<D_{ab}\bigr>\,\bigl<D_{cd}\bigr>\cdot\nn\\
 &~~~\cdot\exp\Bigl[ (\delta_{ac}+\delta_{bd}-\delta_{ad}-\delta_{bc})\,
     \ln(\zeta_*-\zeta'_*) 
     + \bigl< (\varphi_a(\zeta_*)-\varphi_b(\zeta_*))\, 
     (\varphi_c(\zeta'_*)-\varphi_d(\zeta'_*) )\bigr>_{\rm \!c}^{\!(0)}
    \Bigr]. 
\end{align}
We thus identify the annulus amplitude of D-instantons as
\begin{align}
 K_{ab|cd}(z_*,z_*') 
  &=\bigl< (\varphi_a(\zeta_*)-\varphi_b(\zeta_*))\, 
     (\varphi_c(\zeta'_*)-\varphi_d(\zeta'_*) )\bigr>_{\rm \!c}^{\!(0)}
    +(\delta_{ac}+\delta_{bd}-\delta_{ad}-\delta_{bc})\,
     \ln(\zeta_*-\zeta'_*) 
     \nn\\
 &= \bigl< (\phi_a(z_{*})-\phi_b(z_{*})\bigr)\,
    \bigl(\phi_c(z_{*}')-\phi_d(z_{*}')\bigr) 
     \bigr>_{\rm \!c}^{\!\!(0)} \nn\\
 &~~~~+\bigl(\delta_{ac}+\delta_{cd}-\delta_{ad}-\delta_{bc}\bigr)\,
      \ln\Bigl[\sqrt{\mu}\,\bigl(T_p(z_*)-T_p(z_*')\bigr)\Bigr].
\end{align}
The right-hand side can be simplified by using \eq{FZZT-annulus}, 
and we obtain
\begin{align}
 K_{ab|cd}(z_*,z_*')
 &=\ln \frac{(z_{*a}-z_{*c}')(z_{*b}-z_{*d}')} 
          {(z_{*a}-z_{*d}')(z_{*b}-z_{*c}')}\nn\\
 &=\ln \frac{(z_{mn}^--z_{m'n'}^-)(z_{mn}^+-z_{m'n'}^+)} 
          {(z_{mn}^--z_{m'n'}^+)(z_{mn}^+-z_{m'n'}^-)} \nn\\
 &=Z_{\rm annulus}^{(m,n|m',n')}
\end{align}
where we have used the identification (see \eq{zsdzpm}) 
\begin{align}
 z_{*a} = z_{mn}^-,&\quad z_{*b} = z_{mn}^+ , \\
 z_{*c}' = z_{m'n'}^-,&\quad z_{*d}' = z_{m'n'}^+.
\end{align}
This expression correctly reproduces the annulus amplitudes 
of ZZ branes obtained in \cite{Mar,KOPSS}.

\section{Conclusion and discussions}

In this paper, we have studied D-instantons of unitary $(p,p+1)$
minimal strings in terms of the continuum string field theory \cite{fy1}.
In this formulation, there are two types of operators 
that have definite  physical meanings; 
One is the (unmarked) macroscopic loop operator $\varphi_0(\zeta)$, 
and the other is of the soliton operators $D_{ab}$.
We have calculated the expectation value of the soliton operator
$\vev{D_{ab}}$ in the weak coupling limit $g \to 0$.
$\vev{D_{ab}}$ is then expanded as
$\vev{D_{ab}} = \displaystyle\int\!d\zeta\,
\exp[(1/g)\Gamma_{ab} + (1/2)K_{ab}  
+ O(g)],$
 and is dominated by saddle points.
We have carefully identified the valid saddle points, 
and found that $\vev{D_{ab}}$ has a well-defined finite value. 
Since there is no ambiguity in obtaining it, 
this expectation value must be universal.

Each saddle point denoted by $z_*(a,b;l)$ corresponds 
to the location of a localized soliton.
On the other hand, ZZ branes in Liouville field theory
are characterized by a pair of FZZT cosmological constants 
$z_{mn}^\pm$ \cite{Mar}, 
which  correspond to singular points 
on a Riemann surface $\cM_{p,q}$ \cite{SeSh}.
We have shown that the free energy of the localized soliton, 
$\Gamma_{ab}(z_*)$, 
can be identified with minus the partition function of
the ZZ brane $Z_{\rm ZZ}^{(m,n)}$ with the following relation:
\begin{align}
b-a=m,&\quad a+b+l=n ,  \nn \\
z_{*a}(a,b;l) = z_{mn}^-,&\quad z_{*b}(a,b;l) = z_{mn}^+.  
\label{taiou}
\end{align}
This identification is actually valid 
for any $(p,q)$ minimal string  theories.
We pointed out that the set $\bigl\{(a,b;l)\bigr\}$ are redundant 
and can be restricted to a smaller set 
with $1 \leq b-a \,(=m) \leq  p-1$,
$1 \leq a+b+l\,(=n) \leq q-1$ and $mq-np>0$.

With the identification \eq{taiou} 
we have shown that
the annulus amplitudes of localized solitons,  
$K_{ab|cd}(z_*,z_*')$, 
coincide with those of ZZ branes,  
$Z_{\rm annulus}^{(m,n|m',n')}$ 
\cite{Mar,KOPSS}.
There is no ambiguity in deriving this equivalence, 
we thus can conclude that these annulus amplitudes 
also must be universal and can be derived from the saddle point analysis.

For the pure gravity case ($p=2$) 
one can compare the resulting value $\vev{D_{ab}}$ 
with the one-instanton partition function obtained 
with the use of matrix models \cite{david,Hanada:2004im,chemi}.
We found that they coincide up to a single factor $i$. 
In fact, the overall normalization of $D_{ab}$ is not fixed 
only from the continuum loop equations. 
However, if $D_{ab}$ creates a soliton with a definite charge, 
then $D_{ba}$ creates an anti-soliton with the opposite charge. 
Moreover, with the present normalization of $D_{ab}$, the operator 
$D_{ab}\,D_{ba}$ almost gives identity for the 0-soliton sector, 
so that the present normalization seems to be natural. 
In this sense, this discrepancy between our result 
and the matrix model result deserves explanation. 
Among possible ones are 
(i) that it may be natural to introduce $i$ in defining 
the fugacity, 
or (ii) that the D-instanton calculation in matrix models 
may choose a path different 
from the steepest-descent one for $\Gamma_{ab}$.

In this paper, we mainly consider the unitary $(p,p+1)$ minimal strings.
The analysis can be easily extended to general $(p,q)$ minimal strings,
as will be reported in our future communication.

\section*{Acknowledgments}
The authors would like to thank Hikaru Kawai, Ivan Kostov 
and Yoshinori Matsuo for useful discussions. 
This work was supported in part by the Grant-in-Aid for 
the 21st Century COE
``Center for Diversity and Universality in Physics" 
from the Ministry of Education, Culture, Sports, Science 
and Technology (MEXT) of Japan.
MF and SS are also supported by the Grant-in-Aid for 
Scientific Research No.\ 15540269 and No.\ 16740159, 
respectively, from MEXT.

\setlength{\itemsep}{5.\baselineskip}

\end{document}